\documentclass[conference]{IEEEtran}
\IEEEoverridecommandlockouts
\usepackage{tabularx}
\usepackage{subcaption}
\usepackage{cite}
\usepackage{url}
\usepackage{amsmath,amssymb,amsfonts}
\usepackage[numbers]{natbib}
\usepackage{algorithmic}
\usepackage{graphicx}
\usepackage{textcomp}
\usepackage{xcolor}
\usepackage{booktabs}

\usepackage{xspace}
\usepackage{csquotes}
\usepackage{xcolor}
\usepackage{multirow}

\newcommand{\one}{({\em i}\/)\xspace}
\newcommand{\two}{({\em ii}\/)\xspace}
\newcommand{\three}{({\em iii}\/)\xspace}

\def\eg{\emph{e.g.}\xspace}

\def\etal{\emph{et al.}\xspace}
\def\vs{\emph{vs.}\xspace}

\newcommand{\pb}[1]{\vspace{0.75ex}\noindent{\bf \em #1}\hspace*{.3em}}

\def\BibTeX{{\rm B\kern-.05em{\sc i\kern-.025em b}\kern-.08em
    T\kern-.1667em\lower.7ex\hbox{E}\kern-.125emX}}
\begin{document}

\title{Echo Chambers within the Russo-Ukrainian War: The Role of Bipartisan Users\\
}

\author{
\IEEEauthorblockN{
Peixian Zhang, 
Ehsan-Ul Haq, 
Yiming Zhu\textsuperscript{\rm *}, 
Pan Hui\textsuperscript{\rm *}, 
and Gareth Tyson\thanks{\textsuperscript{\rm *}Yiming Zhu is also with The Hong Kong University of Science and Technology, and Pan Hui is also with The Hong Kong University of Science and Technology and University of Helsinki.}}
\IEEEauthorblockA{Hong Kong University of Science and Technology (Guangzhou)}
Email: pzhang041@connect.hkust-gz.edu.cn \quad \{euhaq, yzhucd\}@connect.ust.hk
 \quad \{panhui, gtyson\}@ust.hk
}


\maketitle

\begin{abstract}
 The ongoing Russia-Ukraine war has been extensively discussed on social media. One commonly observed problem in such discussions is the emergence of echo chambers, where users are rarely exposed to opinions outside their worldview. Prior literature on this topic has assumed that such users hold a single consistent view. However, recent work has revealed that complex topics (such as the war) often trigger bipartisanship among certain people. With this in mind, we study the presence of echo chambers on Twitter related to the Russo-Ukrainian war. We measure their presence and identify an important subset of bipartisan users who vary their opinion during the invasion. We explore the role they play in the communications graph and identify features that distinguish them from remaining users. 
We conclude by discussing their importance and how they can improve the quality of discourse surrounding the war.
\end{abstract}

\begin{IEEEkeywords}
Russo-Ukrainian war, Echo chambers
\end{IEEEkeywords}

\section{Introduction}


With the increasing influence of social media on public discourse, individuals' opinions are gaining growing visibility~\cite{misunderstood,naskar2020emotion}.
On the $24^{th}$ February 2022, Russia invaded Ukraine. The invasion has become the subject of significant online debate and has evoked diverse and polarised opinions~\cite{10020274}. Prior research demonstrates that opinions on the conflict vary across regions, with Western Europe and the United States holding different views from those of Eastern Europe and Asian countries~\cite{russia2023}. 
One particular concern is that these trends can result in echo chambers~\cite{garimella2018political,garrett2009echo}, segregating people with opposing stances. 
Echo chambers generally refer to individuals or groups predominantly interacting with those with similar viewpoints, reinforcing and amplifying their existing stances.
This has been shown to create numerous societal issues~\cite{haqWeapon2022}.

Prior literature has analysed users' stances by exploring echo chambers on social media during specific events, \eg the 2018 Brazilian Presidential election~\cite{hyperpartisanship,soares2019asymmetric}, US Presidential election~\cite{longtermpolarization} and COVID-19~\cite{influencer,naskar2020emotion}.  
However, closer inspection reveals that many users do not hold a consistent single stance~\cite{naskar2020emotion, influencer}. 
Such inconsistency in stances impacts users' network centrality and content appreciation~\cite{garimella2018political}. 
Building on this prior work, we aim to explore the presence of echo chambers in online discourse about the Russian invasion of Ukraine. 
Additionally, we investigate whether users display inconsistencies in their polarity as \textit{pro-Russia} or \textit{pro-Ukraine}. 
Specifically, we explore four research questions (RQs):

%



\pb{RQ1:} Do echo chambers exist in Twitter in relation to the Russo-Ukrainian war? 
Considering that echo chambers can have a deleterious impact on the quality of online discourse~\cite{haqWeapon2022}, we seek to measure their presence in the context of the war.

\pb{RQ2:} Are there \textit{bipartisan users} who exhibit both \textit{pro-Russia} and \textit{pro-Ukraine} polarity? \textit{Bipartisan users} are found in various election discussions~\cite {garimella2018political}. 
We explore how such users differ from those with a consistent partisan polarity. 

\pb{RQ3:} Do the \textit{bipartisan users} pay a price of influence? 
Considering that \textit{bipartisan users} are proven to pay a price of content endorsement from others (retweets, likes) on Twitter in some political online discussions~\cite{garimella2018political}, we detect whether the like and retweet influence of \textit{bipartisan users} is less than partisan users in the Russo-Ukrainian war discussion.
If it is different, we then understand how \textit{bipartisan users} interact with the consistent users in the retweet network.


\pb{RQ4:} Do \textit{bipartisan users} help bridge polarised communities and decrease the prominence of echo chambers? 
We study this because partisan users are demonstrated to increase their prominence in the masses by being more partisan~\cite{gromping2014echo}. 

\section{Related Work} \label{sec:related work}



\pb{Polarisation on Social Media:}
Extensive prior work has focused on polarisation, \textit{bipartisan users}, and the properties associated with echo chambers~\cite{haq2020Computational,conover2011political}.
It has been noted that the appearance of echo chambers can lead to several issues that undermine users' communications.
These include rumour cascades~\cite{wang2020viral}, fake news propagation~\cite{sharma2019combating}, and hate speech~\cite{bagavathi2019examining}. 
Moreover, several researchers have observed that political echo chambers emerge on mainstream social media~\cite{gillani2018me}. This could result in a surge of propaganda and partisan content among online communities~\cite{phadke2021makes}. For this, Garrett and Kelly look at how users' news consumption relates to the appearance of echo chambers in online political news sharing~\cite{garrett2009echo}. The authors show that an awareness of the echo chamber's 
early-stage formation can help administrators reduce the spread of extreme ideologies.

\pb{Detection of Echo Chambers:}
There has also been work looking at the automated detection of echo chambers.
Typically, a combination of information sources is used for classification, including textual features (\eg tweet text and hashtags~\cite{baumann2020modeling}), as well as social feedback like retweets, mentions, and followers~\cite{soares2019asymmetric}. 
Other properties can be used to detect echo chambers, \eg~homophily~\cite{cinus2022effect} and social influence~\cite{sasahara2021social}. Some researchers have treated the detection of echo chambers as a network influence estimation problem, leveraging the network structure and subsequently interpreting the centrality metrics to identify communities with echo-chamber characteristics~\cite{garimella2018political, kumar2018community}.
There has also been work exploring the role of online news within echo chambers~\cite{jeon2021chamberbreaker}.
Haq \etal propose a systematic process for rating news articles’ stances to control human bias~\cite{haq2022s}. Their method provides a more objective understanding of news stances, which could facilitate the detection of echo chambers based on news sharing.

\pb{Our Contribution}
In contrast to prior work, we focus on \textit{bipartisan users} in a specific domain: the Russo-Ukrainian war.
Although users who share
inconsistent information are explored in prior work~\cite{garimella2018political,naskar2020emotion,influencer}, this has not focused on echo chambers.
Further, prior work often focuses on a single country \eg the United States.
This means the topic usually pertains to politics and elections. 
In contrast, the global discourse surrounding the Russo-Ukrainian War has not yet been studied in depth.



%


\section{Dataset and Annotation} \label{sec:data}


We now explain the data collection and annotation of tweets to identify \textit{pro-Russia} and \textit{pro-Ukraine} polarity.

\subsection{Tweet Dataset}


We utilise the Twitter dataset on the Russo-Ukrainian war from \cite{haq2022twitter}, which is collected with war-related keywords using Twitter Streaming API.
We utilise a subset of features provided by this dataset -- tweet id, username, tweet text, number of retweets, number of likes, and the ID of referenced tweet (the tweet that is retweeted by current tweet).\footnote{https://developer.twitter.com/en/docs/twitter-api/tweets/lookup/api-reference/get-tweets}

We extract the 16,889,957 (from 337,302 unique users) English language tweets from the dataset. 1.22\% of these are from verified accounts. The tweets cover a period of the first ten days of the war starting from 24$^{th}$ February to 5$^{th}$, March 2022. 
We observe a long-tail distribution of the number of tweets per user
($min=1, max=1858, \mu= 4.38, mid = 1$).
Out of the total tweets, 7.42\% are original tweets, and 78.94\% are retweets, which are the primary focus of our subsequent analysis. 
The remaining tweets consist of quoted and replied tweets.





\subsection{Manual Annotation of Tweets}\label{sec: annotation}

We next label a subset of tweets with their polarity.
In order to reduce the terms in the paper, we use the term polarity in this paper to take the place of stance.
We categorise tweets into one of three polarity labels -- \textit{pro-Russia}, \textit{pro-Ukraine}, or \textit{not-sure}.
To select tweets that mention Russia and Ukraine, we first filter the tweets that contain any relevant keywords or hashtags: ``Russia'', ``Ukraine'', \#IStandwithRussia, \#StopRussia, \#IstandwithPutin and \#RussiaUkraineWar). We then randomly sample 2,205 tweets to manually annotate their polarity in text.
However, we observe that not all tweets directly state their polarity towards Russia or Ukraine. Instead, they might indicate their leaning based on other entities, such as politicians, countries, and regions. For instance, here is a tweet indicating its \textit{pro-Russia} stance:
\begin{displayquote}
\textit{``Dear my president, president Vladimir Putin, keep strong protect your nation against the evil NATO and America! \#IstandwithPutin''}
\end{displayquote}
To capture these nuances, in annotation, we also consider users' stance (``pro'' or ``anti'') towards relevant entities mentioned in the text. For example, ``pro Putin'', ``anti nato'', and ``anti US'' are assigned to the above example tweet, and our final decision for its stance is \textit{pro-Russia}.
We use the following entity-related attitudes to annotate tweets' polarity:

\begin{itemize}
    \item \textbf{\textit{Pro-Russia}:} pro Putin, anti Biden, anti US, anti Trump, anti Lukashenko, anti Carlson, pro Russia(n), anti Kamala, anti Ukraine, anti Ukrainian, anti Zelensky, anti NATO, anti GOP (Grand Old Party), anti Democrats, anti West
    
    \item \textbf{\textit{Pro-Ukraine}}: pro US, pro Zelensky, pro Ukraine, pro Ukrainian, pro Biden, anti Putin, anti Russia(n), anti Oligarch, anti Belarus, pro Trump.
\end{itemize}



Overall, the manually annotated data includes 539 tweets labelled as \textit{pro-Russia}, 938 tweets labelled as \textit{pro-Ukraine}, and 728 tweets labelled as \textit{not-sure}. 






\section{Quantifying Polarity} \label{sec:method}

We next use our annotated dataset to train a classifier that predicts the polarity of the rest of the tweets in the dataset. 



\subsection{Predicting Tweet Polarity} \label{sec:classifier} 




We use five commonly used text embedding methods -- BERT \cite{bert}, Sentence Transformer \cite{sentencetran}, Universal Sentence Encoder \cite{use_encodingl}, Word2Vec \cite{word2vec}, and TF-IDF~\cite{manning2009introduction} to train several classifiers to select the best performing one. We preprocess tweets' text before applying the embedding methods. Our process involves:
\one Removing any mentions for users (\texttt{@\{username\}}) or retweeting (\texttt{RT@\{username\}}). \two Removing all hyperlinks and emojis. \three Removing all non-alphanumeric characters, including punctuation and special symbols (\eg \#, @, \$).

We then use six commonly used machine learning algorithms (SVM, KNN, Decision Tree, Random Forest, Naive Bayes, and Logistic Regression)~\cite{scikit-learn} and train multiple classifiers using the above-mentioned text features.
We use grid search for each combination of an embedding and an algorithm to identify the best hyper-parameters~\cite{scikit-learn}.
We evaluate each combination using 5-fold cross-validation. Table~\ref{tab: embedding} summarises the best combination for each embedding and the corresponding F1-Score. 
The classifier with Sentence Transformer and SVM combination achieves the highest performance (F1-Score $=0.70$). Finally, we select this classifier to predict the rest of the tweets' polarity in our dataset. The classifier outputs probabilities for the three labels (\textit{pro-Russia}, \textit{not-sure}, \textit{pro-Ukraine}), where each probability represents the likelihood of a tweet leaning towards the corresponding polarity. 

\begin{table}[t!]
\centering
\begin{tabular}{llc}
\toprule
\textbf{Embedding}            & \textbf{Algorithm}    & \textbf{F1-Score} \\ \midrule
TF-IDF                        & NaiveBayes        & 0.63                    \\
Bert                          & SVM               & 0.65                    \\
Word2Vec                      & LogisicRegression & 0.59                    \\
Sentence Transformer & SVM      & \textbf{0.70}            \\
Universal Sentence Encoder    & SVM               & 0.68                    \\ \bottomrule
\end{tabular}
\caption{The list of five embedding methods corresponding to the best performance classifier models. The bold F1-Score (0.70) denote that the combination of Sentence Transformer and SVM achieves the highest performance in our experiment.}
\label{tab: embedding}
\end{table}

\subsection{Quantifying Users' Polarity} 

We then quantify users' polarity based on their tweets' polarity. We first encode tweets' polarity into numerical notations: \textit{pro-Ukraine} $=1$, \textit{not-sure} $=0$, and \textit{pro-Russia} $=-1$. Given a tweet $t$, we quantify its polarity $s_t$ as: $s_t= \sum_{l} p_l *l$, where $l$ denotes the polarity label ($l\in[-1,0,1]$) and $p_l$ denotes the probability output by the classifier for polarity $l$. Following this, for a user $u$, we quantify the user's polarity $g_u$ as the average of the tweets' polarity posted by this user:\[g_u = \frac{\sum_{i=1}^{n}s_{t_i}}{n}\]where $t_i(i\in[1,n])$ denotes the tweet posted by $u$.

Accordingly, when $g_u$ is close to 1, it indicates that the user is \textit{pro-Ukraine}, and a score close to -1 indicates that the user is more \textit{pro-Russia}. The polarity distribution of all users ranges from -0.99 to 0.97, with a mean of 0.27 and a median of 0.288.

\section{Results \& Analysis} \label{sec: analysis}

The above dataset and classifier allow us to estimate the polarity of each user.
We next exploit our data to answer our research questions.

\subsection{RQ1: Do echo chambers exist in Twitter in relation to the Russo-Ukrainian war?}
\label{sec: echo chamber}

To detect the echo chambers, we utilise the retweet network following the approach proposed in~\cite{cinelli2021echo}. We use two measures \one homophily with neighbours
and \two homophily in communities.


\pb{Detecting Echo Chambers by Homophily:}
We construct the retweet network as a weighted directed graph, where edge $(i,j)$ is directed from user $i$ to $j$, if $i$ has been retweeted by $j$, and the weight is the number of retweets by $j$ towards $i$.
We focus on the active users who have produced multiple retweets (weighted in-degree $\geq2$) or been retweeted by other users several times (weighted out-degree $\geq2$).
This is because 53.39\% of users in the network have only produced one retweet and we set $\geq2$ threshold to reduce noise and network sparsity. In all, the retweet network contains 1,488,984 nodes and 8,475,794 edges.

The first definition of echo chambers we use is based on users' homophily with their neighbours.
Individuals tend to adopt the same polarity as their friends, limiting their access to diverse information sources. 
This homophily phenomenon contributes to the creation of echo chambers~\cite{cinelli2021echo}. Thus, we assess the existence of echo chambers by measuring the homophily of each user with their neighbours in the retweet network.
We follow the approach in~\cite{cinelli2021echo} and calculate the average polarity of retweets from neighbourhoods
to assess homophily between nodes and their neighbourhoods.
Given a node (user, with at least two neighbour) $u$ with a polarity $g_u$, its average polarity of retweets from neighbourhood is calculated as $u_i^N=\frac{1}{k_i}\sum_{j} A_{ij}u_j$ ($i \neq j$ to avoid self-retweeting), where $A$ is the adjacency matrix of the retweet network.
$A_{ij}=weight$ if edge $(j,i)$ exists; otherwise,
$A_{ij}=0$.
$k_i=\sum_{j}A_{ij}$ is the weighted in-degree of node $i$.

\begin{figure}[t]
     \centering
     \begin{subfigure}[b]{0.20\textwidth}
         \centering            \includegraphics[width=\textwidth]{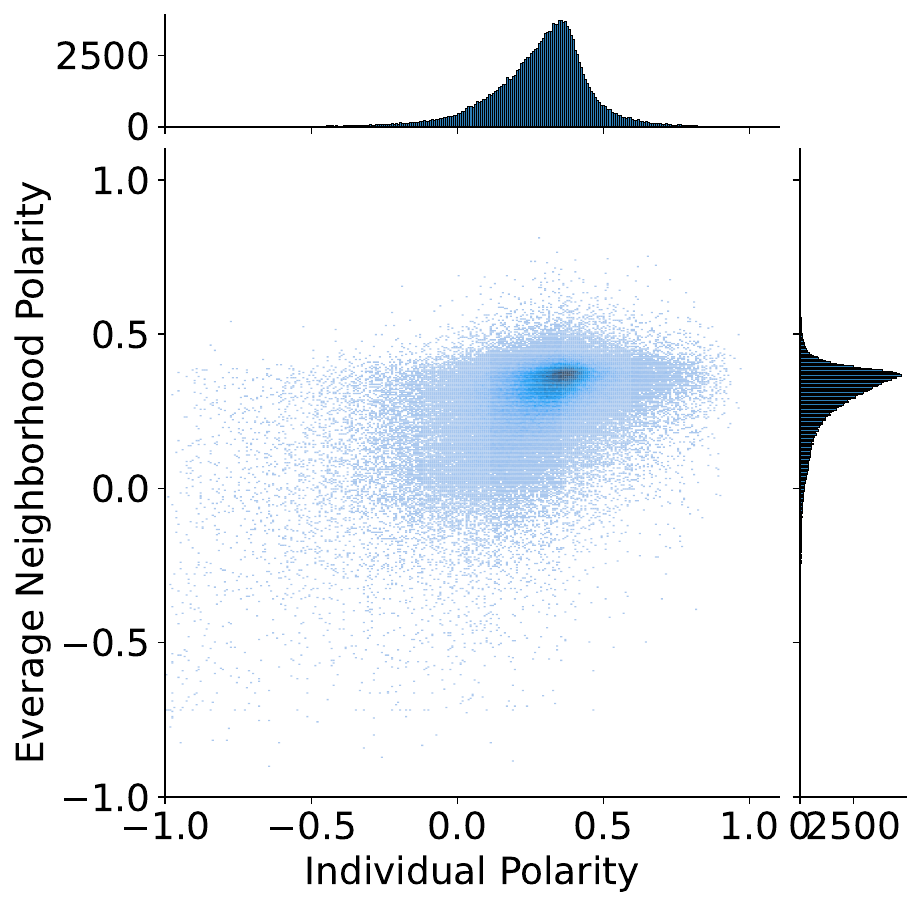}
            \caption{Neighbours}
            \label{fig:echo_chamber_p_2}
     \end{subfigure}
     \begin{subfigure}[b]{0.245\textwidth}
         \centering        \includegraphics[width=\textwidth]{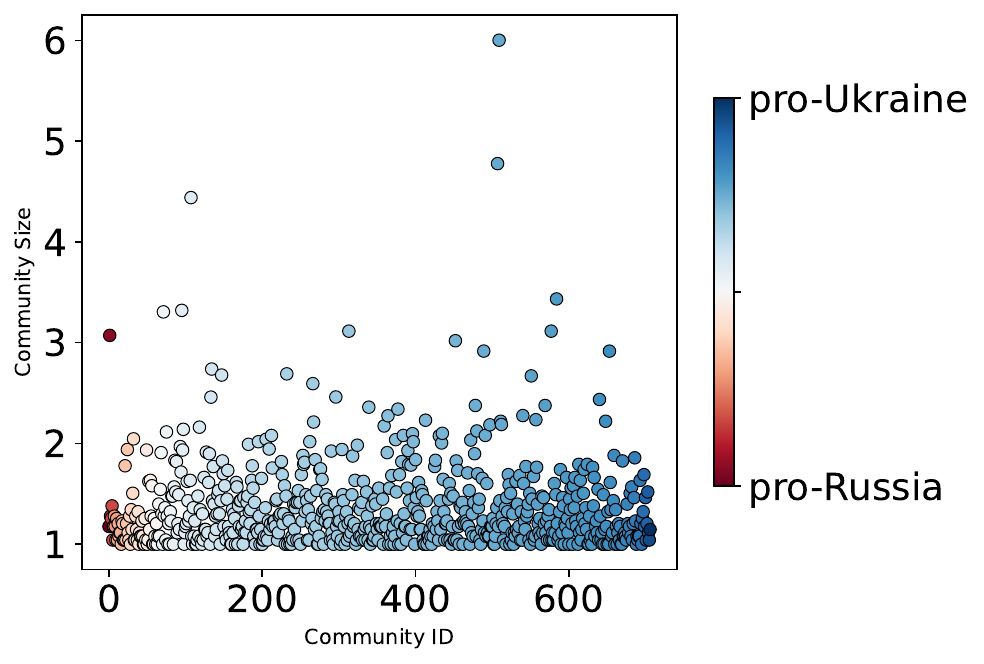}
        \caption{Communities}
        \label{fig:echo_chamber_h}
        \end{subfigure}
        \caption{a) Joint distribution of the individual polarity and the average of their neighbourhood polarity for the dataset. The marginal distribution of individual polarity and average neighbourhood polarity are plotted on the $x-axes$ and $y-axes$, respectively. b) Size and average polarity of communities detected by Louvain algorithm with resolution 0.1. Each community has more than ten nodes and $x$-axis sorted by the polarity of communities.}
        \label{fig:rq1_figure}
\end{figure}

The second definition of echo chambers is based on users’ homophily in communities. 
The homophily with neighbours provides an intuitive understanding of the echo chamber.
The neighbours of users play a significant role in determining the type of content they are likely to engage with.
When a user's neighbours have similar polarity, there is a higher likelihood that the user be exposed to content that aligns with their own polarity.
However, it does not provide a higher level of view to understand the interactive activities.
Then we also take advantage of the community detection in the retweet network to understand the size of the user cluster.
To measure the homophily of the community, we extract communities in the retweet network using the Louvain algorithm~\cite{Louvain} with resolution 0.1 and remove the communities with fewer than ten nodes(users), leaving 707 communities.
The resulting communities obtained from the Louvain algorithm represent clusters of nodes that exhibit strong interconnectivity and share similar patterns of connections~\cite{Louvain}.
Then we compute the average polarity of each community determined as the average individual polarity of the users in the community~\cite{cinelli2021echo}. 

\pb{Results:}
The results of the two methods show in Figure~\ref{fig:echo_chamber_p_2} and~\ref{fig:echo_chamber_h}.
Figure~\ref{fig:echo_chamber_p_2} shows the individual polarity \vs the average polarity of retweets from neighbours, where a darker area means a higher density of users on distribution. 
We observe strong clustering patterns of users' polarity on the diagonal of the plot, indicating a positive correlation between these two metrics. A Pearson correlation test between individual and neighbour polarity supports this conclusion $(r=0.49, p<0.001)$.
Such a result implies that some users are likelier to retweet others with a similar polarity.
This indicates the homophily by users' polarity on the retweet network. 
Moreover, according to~\cite{cinelli2021echo}, the strong clustering patterns of users' polarity with the appearance of homophily evidences the existence of an echo chamber.
Figure~\ref{fig:echo_chamber_h} shows the communities arranged by increasing average polarity on the $x$-axis from blue to red, and the $y$-axis refers to the community size processed by log10. 
The communities span the whole colour spectrum and form some communities with average polarity. 
There are relatively few communities around the 0 territories, indicating that few communities are entirely neutral. 
Furthermore, we observe that the communities with a polarity close to -1 tend to be relatively smaller compared to communities with a polarity close to 1.


Based on the analysis of the neighbouring homophily of the retweet network in Figure~\ref{fig:echo_chamber_p_2}, we observe a clear echo chamber in \textit{pro-Ukraine} polarity with a distinct cluster of users have similar polarity with their neighbours.
Individuals who exhibit polarity towards \textit{pro-Ukraine} polarity tend to reside near most like-minded individuals within their community, engaging in frequent interactions with them. 
Then, based on the community structure in Figure~\ref{fig:echo_chamber_h}, we observe the presence of communities with varying polarity between \textit{pro-Russia} and \textit{pro-Ukraine}.
The discourse observed on the Twitter platform demonstrates the existence of an echo chamber on both \textit{pro-Ukraine} and \textit{pro-Russia} and is characterised by the prevalence of information favouring a \textit{pro-Ukraine} perspective.



\begin{table}[t!]
\centering
\resizebox{.85\columnwidth}{!}{
\begin{tabular}{lllll}
\toprule
\textbf{Groups}                          & \textbf{Tweets}   & \textbf{Users}   & \multicolumn{2}{l}{\textbf{Verified Users}} \\ \midrule
\textbf{\textit{pro Ukraine}} & 1,3177,944 & 1,276,671 & 20,941 &(1.64\%)          \\
\textbf{\textit{pro Russia}}  & 191,208    & 65,380    & 401 &(0.61\%)             \\ \midrule
\textbf{Bipartisan Users}        & 616,074    & 130,170   & 853 &(0.66\%)            \\ \bottomrule
\end{tabular}
}
\caption{Distribution of the users and their generated content into partisan (\textit{pro Ukraine} and \textit{pro Russia}) and \textit{bipartisan users}. The percentages denote the proportions of verified users to the total users base in each group.}
\label{tab: groups}
\end{table}


\subsection{RQ2: Are there \textit{bipartisan users} who exhibit both \textit{pro-Russia} and \textit{pro-Ukraine} polarity? } \label{sec:bipartisan users}


We detect the users sharing tweets 
includes both \textit{pro-Ukraine} and \textit{pro-Russia} by the classifier and category users by their tweets. 

\pb{Identifying Bipartisan Users:}
We utilise the labels obtained from the classifier to categorise the users, which is easier to process when dealing with categorical data compared to continual data.
We focus on users who have submitted multiple tweets ($>1$), as the users with only one single tweet are not deemed as holding a consistent attitude.
We then classify the users by their tweets.
Notably, to avoid confusion about the tweet's polarity and user categories in similar terms, the term without a hyphen(-) points to users in the corresponding group and the term has a hyphen(-) for the rest of the situations.
For analysis, we classify the users into two categories below:

\begin{itemize}
    \item \textbf{Bipartisan Users:} {A user is classified as \textit{bipartisan user} if user has at least 20\% \textit{pro-Russia} and at least 20\% \textit{pro-Ukraine} tweets. 20\% is the minimum thresholds to distinguish \textit{bipartisan users} and partisan users in several topics~\cite{garimella2018political}.}



    \item\textbf{Partisan Users: }{If only \textit{pro-Russia} or \textit{pro-Ukraine} tweets cover more than 20\% of a user's total tweets individually, this user is classified as a partisan user. According to the major tweets' polarity, a partisan user is further categorised into \textit{pro Russia} or \textit{pro Ukraine}.}


    \item \textbf{Not Sure:} {If a user is neither a \textit{bipartisan user} nor a partisan user, this user is classified as \textit{not sure}.}
\end{itemize}




\begin{figure*}[t!]
     \centering
     \subfloat[\textit{bipartisan Users}\label{fig: inconsist}]{
         \centering
         \includegraphics[width=0.23\textwidth]{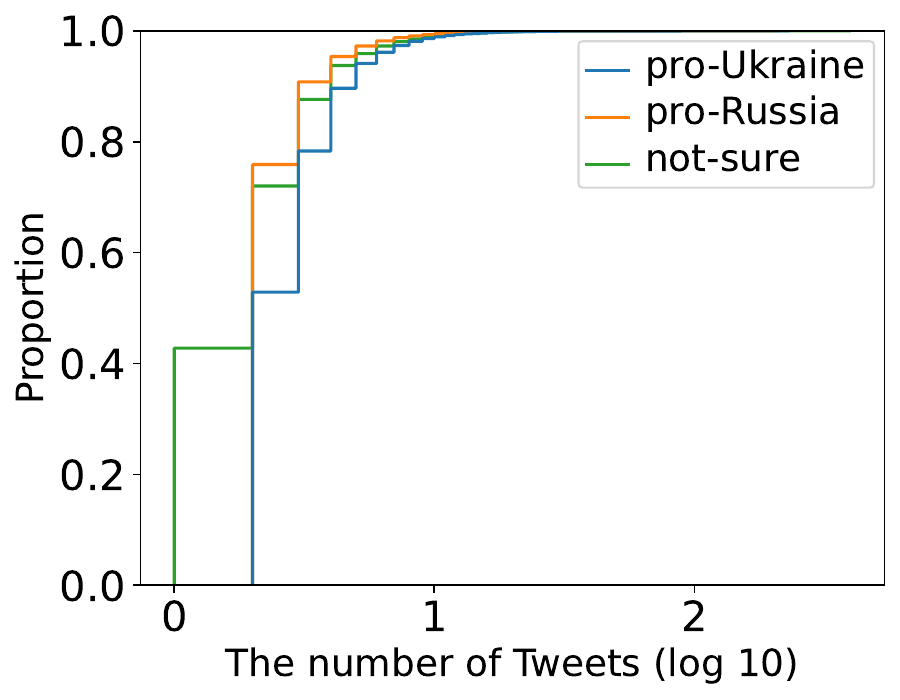}
         
    }  
     \subfloat[\textit{pro Ukraine}]{
         \centering
         \includegraphics[width=0.23\textwidth]{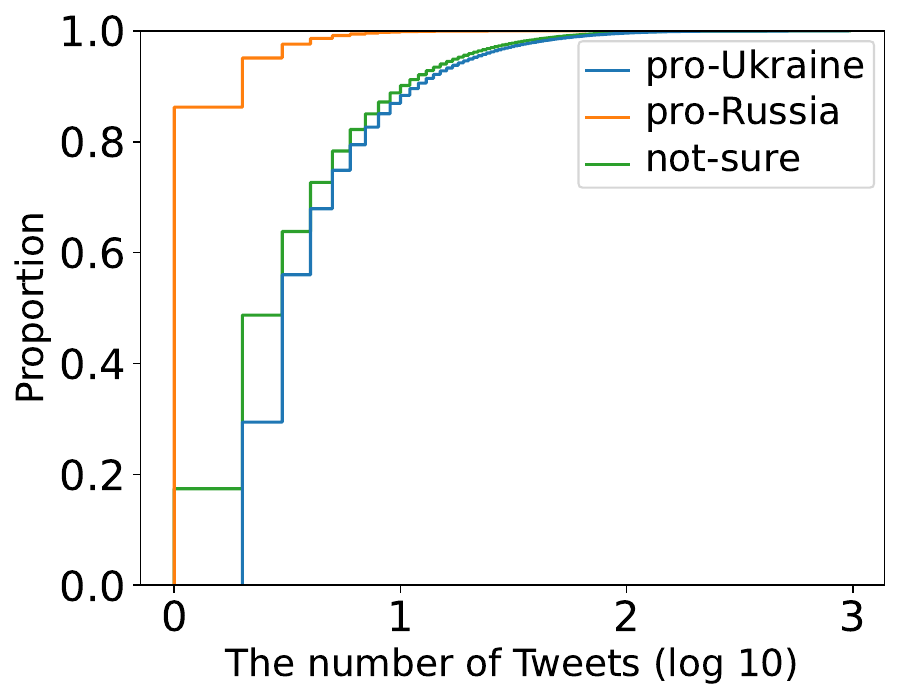}
         \label{fig: pro_ukraine}
     }
     \subfloat[\textit{pro Russia}]{
         \centering
         \includegraphics[width=0.23\textwidth]{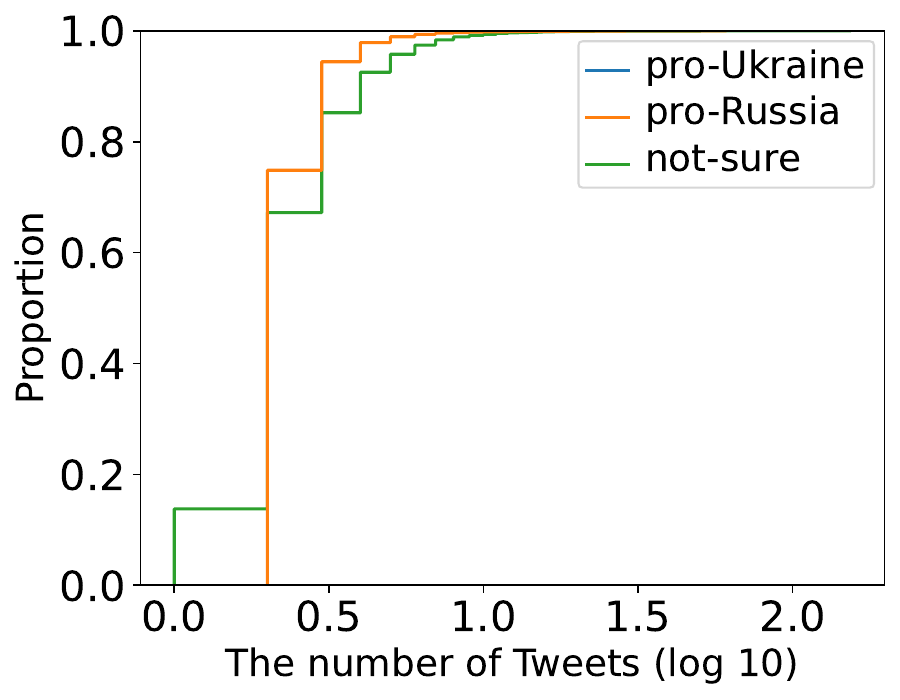}
         \label{fig: pro_russia}
     }  
     \caption{Display the cumulative distribution function for the number of polarity of tweets in the three groups, respectively. The $x$-axis shows the logarithm base 10 of the number of tweets, where each value is incremented by 1 to mitigate the impact of zero.}
     \label{fig:stance}
\end{figure*}

\pb{Results:} 
We observe users exhibiting bipartisan polarity, including both \textit{pro-Russia} and \textit{pro-Ukraine} tweets. 
Our dataset contains 130,170 \textit{bipartisan users}, 65,380 \textit{pro Russia} users and 1,276,671 \textit{pro Ukraine} users. 
Table~\ref{tab: groups} presents the distribution of users' polarity. We note that \textit{bipartisan users} are the second largest in the dataset, covering 8.84\% users and 4.41\% tweets.
In addition, we also observe a strong bias towards \textit{pro Ukraine} among Twitter users, where \textit{pro Ukraine} group consists of 86.72\% users and 94.23\% tweets. 
\textit{Pro Russia} is a minor class, constituting only 4.44\% users and 1.37\% of tweets. 
Moreover, we find that \textit{pro Ukraine} group (1.64\%) contains more verified users than \textit{pro Russia} (0.61\%) and \textit{bipartisan users} (0.66\%).
After detecting \textit{bipartisan users},
we then inspect the difference in a tweet posting behaviour among different user groups. 
Thus, we examine the distribution of tweets' polarity within each group. 
Figure~\ref{fig:stance} displays the cumulative distribution function (CDF) for the number of tweets with polarity in the three groups. 
The $x$-axis shows the logarithm base 10 of the number of tweets$+1$. 
Figure~\ref{fig: inconsist} illustrates that for \textit{bipartisan users}, \textit{pro-Ukraine} tweets have the highest frequency, followed by \textit{not-sure} tweets, while \textit{pro-Russia} tweets have the lowest frequency.
Moreover, we find that the distribution of tweets' polarity in \textit{bipartisan users} appears to be distinct from those in other groups by posting relatively more \textit{not-sure} tweets (25.71\%).
Interestingly, we also notice in Figure~\ref{fig: pro_ukraine} that 13.76\% of \textit{pro Ukraine} users have \textit{pro-Russia} tweets.
However, none of \textit{pro Russia} users has ever posted \textit{pro-Ukraine} tweets in Figure~\ref{fig: pro_russia}.

We character \textit{bipartisan users} tweet in both \textit{pro-Ukraine} and \textit{pro-Russia}.
Furthermore, based on our dataset, while some \textit{pro Ukraine} users share tweets with opposing polarity, none of the \textit{pro Russia} users exhibit similar activities.

\subsection{RQ3: Do \textit{bipartisan users} pay a price in terms of influence? 
How do \textit{bipartisan users} interact with the consistent users in the retweet network?} \label{sec:rq3}






We now assess users' influence by the number of retweets and likes and characterise \textit{bipartisan users}.
We analyse the influence of each group individually. Then, we understand the users and content retweets across \textit{bipartisan users} and partisan users.



\pb{Bipartisan Users' Influence:} 
We aim to analyse whether \textit{bipartisan users} pay a price in terms of influence.
We follow the interpretation
of users' influence in~\cite{cha2010measuring} taking follow counts and like counts as proxy and focus on those users with multiple followers in our dataset, where we have also excluded 24,122 ($1.64\%$ of the total) accounts without any followers. 
On this basis, we count the average of retweets and likes received per tweet normalised on user's followers count to measure user's influence. 
A higher normalised average number of retweets (likes) indicates the user has a higher influence.
Then, we aim to further understand \textit{bipartisan users}' retweeting interaction with \textit{pro Russia} and \textit{pro Ukraine} users, including users and content. 
Thus, we focus on retweetees (users who get retweeted) and retweeters (users who get retweeted) connected to \textit{bipartisan users}.
In the retweet network, a ``predecessor'' node denotes a retweetee and a
``successor'' node denotes a retweeter. 
We show that partisan users connect with \textit{bipartisan users} in retweet network and analyse the polarity of users from \textit{pro Russia} and \textit{pro Ukraine} groups to see whether existing distinctive features in these groups.
After that, we analyse the content partisan users get from \textit{bipartisan users}.

\begin{figure}[t!]
    \centering
\subfloat[Retweet Count\label{fig:cdf_retweet_sub}]{
    \includegraphics[width = 0.23\textwidth]{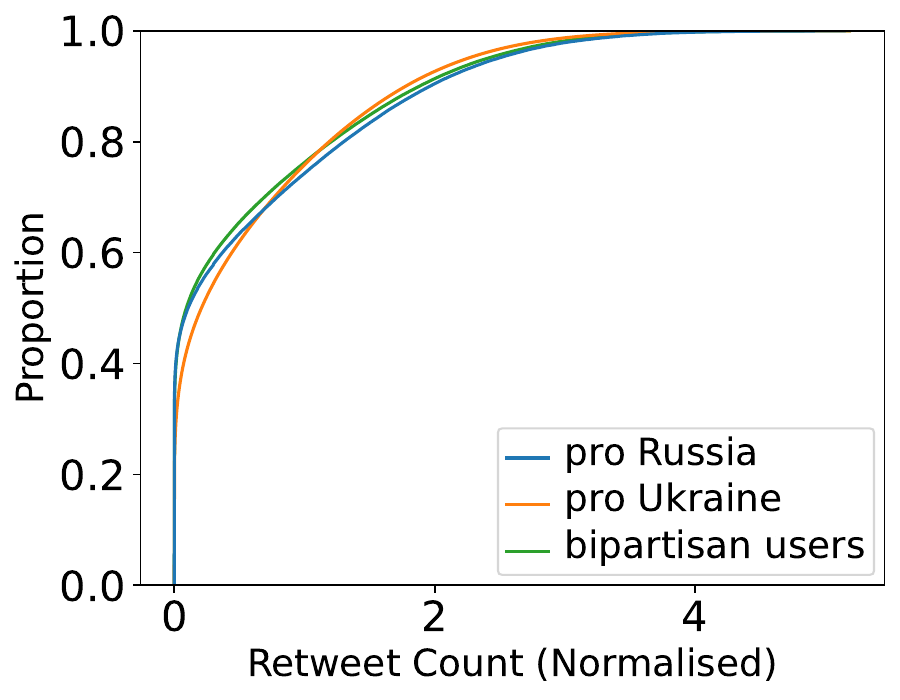}
}
\subfloat[Like Count\label{fig:cdf_like}]{
    \includegraphics[width = 0.23\textwidth]{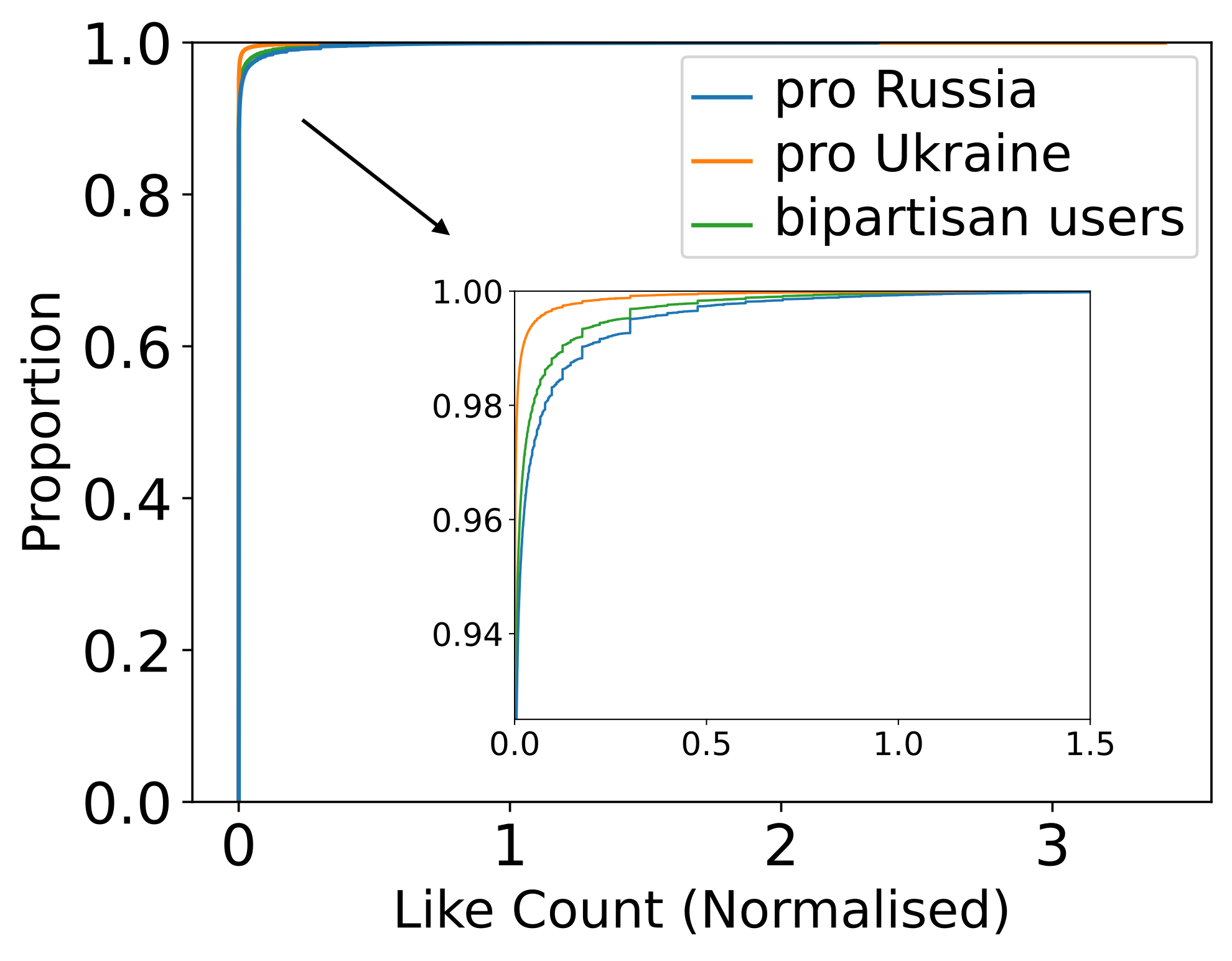}
     
} 
    \caption{Cumulative distribution function (CDF) shows the retweet counts and like counts normalised by followers in \textit{pro Russia}, \textit{pro Ukraine} and bipartisan group. }
    \label{fig:cdf_influence} 
\end{figure}

\begin{table*}[t!]
\scriptsize
\centering
\begin{tabular}{llll}
\toprule
\textbf{Metrics}                          & \textbf{KW H(2)} & \textbf{Mean diff (post-hoc)}                                & \textbf{$p-value$(post-hoc)} \\ \hline
\multirow{3}{*}{\begin{tabular}[c]{@{}l@{}}Normalised retweet count\\ (Figure~\ref{fig:cdf_retweet_sub})\end{tabular}} & 23081.417        & \textit{pro Ukraine}(78.939) \textless \textit{pro Russia}(146.343)            & ***                        \\
                                          & (***)            & \textit{pro Ukraine}(78.939) \textless \textit{bipartisan users}(115.560)      & ***                        \\
                                          &                  & \textit{pro Russia}(146.343) \textgreater \textit{bipartisan users}(115.560)   & ***                        \\ \hline
\multirow{3}{*}{\begin{tabular}[c]{@{}l@{}}Normalised like count\\(Figure~\ref{fig:cdf_like})\end{tabular}}    & 31217.430        & \textit{pro Ukraine}(0.001) \textless \textit{pro Russia}(0.044)               & ***                        \\
                                          & (***)            & \textit{pro Ukraine}(0.001) \textless \textit{bipartisan users}(0.027)         & ***                        \\
                                          &                  & \textit{pro Russia}(0.044) \textgreater \textit{bipartisan users}(0.027)       & ***                        \\ \hline
\multirow{3}{*}{\begin{tabular}[c]{@{}l@{}}Polarity in pro Russia\\(Figure~\ref{fig:prosuc_russia})\end{tabular}}    & 1246.980         & entire group(-0.193) \textgreater successor nodes(-0.270)    & ***                        \\
                                          & (***)            & entire group(-0.193) \textgreater predecessor nodes(-0.307)  & ***                        \\
                                          &                  & successor nodes(-0.270) \textless predecessor nodes(-0.307)  & ***                        \\ \hline
\multirow{3}{*}{\begin{tabular}[c]{@{}l@{}}Polarity in pro Ukraine\\(Figure~\ref{fig:prosuc_ukraine})\end{tabular}}  & 10497.530        & entire group(0.357) \textgreater successor nodes(0.319)      & ***                        \\
                                          & (***)            & entire group(0.357) \textgreater predecessor nodes(0.313)    & ***                        \\
                                          &                  & successor nodes(0.319) \textgreater predecessor nodes(0.313) & ***                        \\ \bottomrule
\end{tabular}
\caption{Pair-wise comparison of groups by the
Kruskal Wallis test with Dunn’s post-hoc test corresponding to different figures. ``Mean diff" column shows the comparison results of the mean value of corresponding metrics between three groups. *** denotes that $p < 0.001$.}
\label{tab: sta_result}
\end{table*}
\pb{Results:}
Figure~\ref{fig:cdf_retweet_sub} and~\ref{fig:cdf_like} display the cumulative distribution function for retweets and likes on the tweets, respectively. 
The $x$-axis represents the log of $1 + $ the average number of retweets (likes), and the $y$-axis represents the cumulative probability. 
Figure~\ref{fig:cdf_retweet_sub} shows that different retweet distributions exist across the three groups. 
Figure~\ref{fig:cdf_like} shows that nearly 80\% of users received no likes for their tweets. 
\textit{Pro Ukraine} group gains the fewest likes,
followed by \textit{bipartisan users}.
\textit{Pro Russia} users get the most likes of all. 
Table~\ref{tab: sta_result} shows statistics details of these distributions and the statistical significance of these distributions with the Kruskal-Wallis (KW) test and Dunn's test, indicating that the influence of \textit{bipartisan users} is not reduced compared to partisan users. 


\begin{table}[t!]
\scriptsize
\centering
\begin{tabular}{llll}
\toprule
                          & \textbf{\textit{pro Ukraine}} & \textbf{\textit{pro Russia}} & \textbf{\textit{Bipartisan}} \\ \midrule
\textbf{Predecessor Nodes} & 2.98\%               & 5.96\%              & 4.59\%              \\
\textbf{Successor Nodes}  & 11.98\%              & 21.71\%             & 23.06\%             \\ \bottomrule
\end{tabular}
\caption{The table presents the proportional distribution of predecessor and successor nodes connected to bipartisan nodes, with the values normalised by the total number of nodes within each respective group.}
\label{tab: pre_suc nodes}
\end{table}

\begin{table}[t!]
\scriptsize
\centering
\begin{tabular}{llll}
\toprule
                              & \textit{\textbf{pro-Ukraine}} & \textit{\textbf{pro-Russia}} & \textit{\textbf{not-sure}} \\ \midrule
\textit{\textbf{pro Ukraine}} & 62.66\%                       & 9.67\%                       & 27.67\%           \\
\textit{\textbf{pro Russia}}  & 0.00\%                        & 75.56\%                      & 24.44\%           \\ \bottomrule
\end{tabular}
\caption{The table presents the percentage of the tweets successor nodes retweet from \textit{bipartisan users} normalised by the total number of groups, respectively. }
\label{tab: suc_tweets}
\end{table}



\begin{figure}[t]
     \centering
     \begin{subfigure}[b]{0.22\textwidth}
         \centering
            \includegraphics[width=\textwidth]{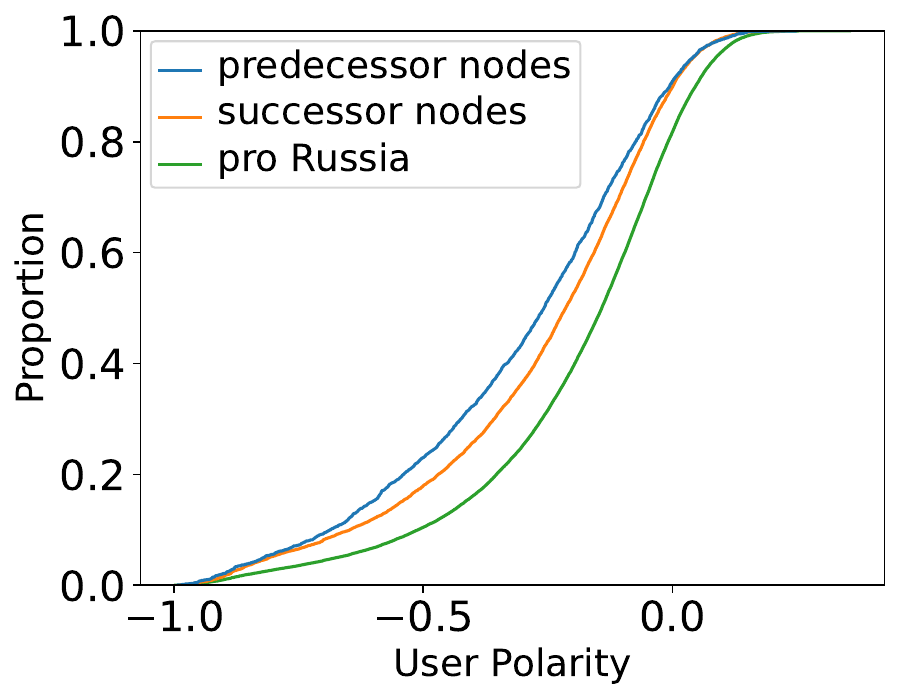}
            \caption{pro Russia}
            \label{fig:prosuc_russia}
     \end{subfigure}
     \begin{subfigure}[b]{0.22\textwidth}
         \centering
        \includegraphics[width=\textwidth]{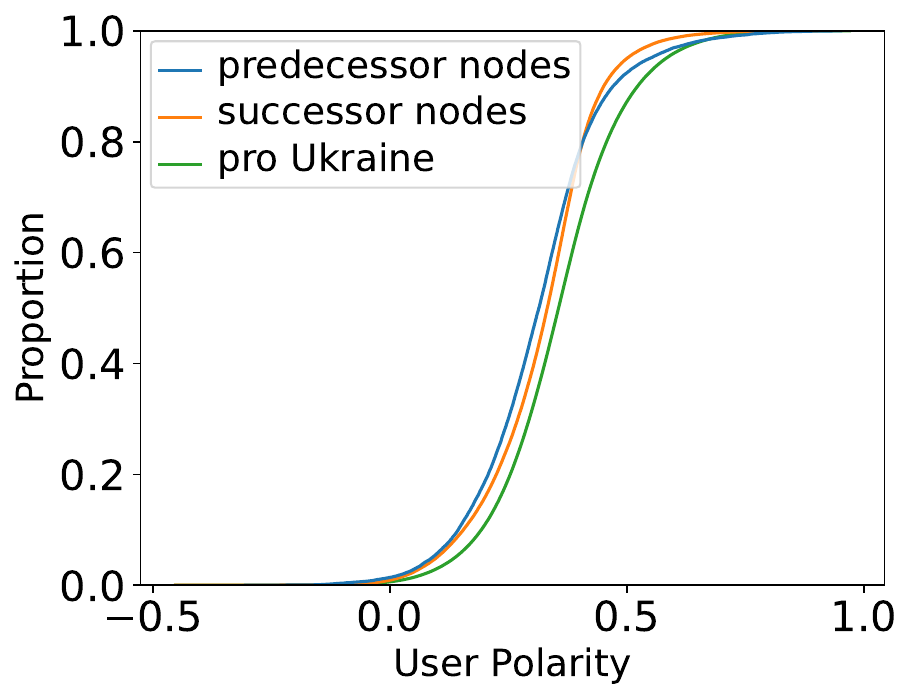}
                    \caption{pro Ukraine}
        \label{fig:prosuc_ukraine}
        \end{subfigure}
        \caption{a) The cumulative distribution function for users polarity of \textit{pro Russia} groups and predecessor nodes and successor nodes of bipartisan in \textit{pro Russia} groups. b) The cumulative distribution function for users' polarity of \textit{pro Ukraine} groups and predecessor nodes and successor nodes of bipartisan in \textit{pro Ukraine} groups.}
        \label{fig:rq3_figure_2}
\end{figure}

Table~\ref{tab: pre_suc nodes} presents the distribution of polarity in predecessor and successor nodes connected to \textit{bipartisan users}, where the percentages show the proportion of nodes with certain polarity to the total node base in the corresponding group.
Notably, only a small percentage of users appear in both predecessor and successor categories; 2.00\% in \textit{pro Russia} and 3.17\% in \textit{pro Ukraine}.
Figure~\ref{fig:prosuc_russia} and Figure~\ref{fig:prosuc_ukraine} show the cumulative distribution of users' polarity \textit{pro Russia} and \textit{pro Ukraine} groups, respectively. 
Within each figure, the green line is the polarity for all users in the respective group, whereas the Orange and Blue lines show the polarity of users categorised as successors and predecessors in their interactions with \textit{bipartisan users}, respectively.
Figure~\ref{fig:prosuc_russia} shows that in \textit{pro Russia} group, the individuals interacting with \textit{bipartisan users} belong to distinct subgroups with more polarisation polarity close to 1 and -1.  
For \textit{pro Russia} group, the entire group exists polarity distribution 
that is higher than successor nodes 
and predecessor nodes.
Notably, when the polarity is close to -1, it signifies extreme polarity.
Conversely, Figure~\ref{fig:prosuc_ukraine} shows that in \textit{pro Ukraine} group, the phenomenon is the opposite. 
For \textit{pro Ukraine} group, the whole group has larger polarity distribution 
than successor nodes 
and predecessor nodes.
Similarly, Table~\ref{tab: sta_result} summarises statistics details of these distributions and statistical significance of pairwise comparison test.
Table~\ref{tab: suc_tweets} presents the percentage of the tweets successor nodes retweet from \textit{bipartisan users}. 
When \textit{pro Ukraine} and \textit{pro Russia} users retweet content from \textit{bipartisan users}, they demonstrate a tendency to select tweets that align with their respective positions. Specifically, \textit{pro Ukraine} users exhibit \textit{pro-Ukraine} polarity in 62.22\% of the retweeted tweets, while \textit{pro Russia} users retweet content from \textit{bipartisan users} with \textit{pro-Russia} polarity in 75.56\% of the retweeted tweets.

Based on the analysis result, \textit{bipartisan users} do not show the lowest retweet and like counts among groups.
Moreover, amongst \textit{pro Russia} users who retweet content from \textit{bipartisan users}, polarity close to -1 is observed compared to the overall group.
Conversely, for \textit{pro Ukraine} users, the opposite situation is observed, where there is less polarisation among those who retweet from \textit{bipartisan users}.
Furthermore, both \textit{pro Ukraine} and \textit{pro Russia} users interact with \textit{bipartisan users} and show a clear preference for content along with their polarity.


\subsection{RQ4: Do \textit{bipartisan users} help bridge polarised communities, and decrease the prominence of echo chambers?} \label{sec:rq4}


We next investigate if \textit{bipartisan users} mitigate the prominence of echo chambers by removal them from the largest community. 


\pb{Bipartisan Users Removal from the Community:}
We take the largest community detected by Louvain algorithm with a resolution of 0.1.
containing 68.30\% of the nodes in the network of the retweet network in Figure~\ref{fig:echo_chamber_h}.
The largest community consists of 79.15\% \textit{pro Ukraine}, 2.68\% \textit{pro Ukraine}, 6.96\% \textit{bipartisan users}, and the remaining portion comprising \textit{not sure} users.  
For a baseline comparison, we first select a control group consisting of non-bipartisan users. For a given user in \textit{bipartisan users}, we take the 
degree of the user (node) and find another user in the non-bipartisan users who has the closest matching degree.
Out of all \textit{bipartisan users}, only 11 (0.01\%) users cannot find non-bipartisan users who have the same degree.
The largest difference in degree between a \textit{bipartisan user} and a non-bipartisan user is 25 degrees, where the bipartisan node has a degree of 4504.
We use these non-bipartisan users for comparison against \textit{bipartisan users} to characterise the role of the latter in polarised communities.
According to Section~\ref{sec: echo chamber}, we take communities with absolute polarity larger than 0.5 as polarised communities, which suggests evidence of echo chambers.
In order to understand whether \textit{bipartisan users} help bridge polarised communities, we next take the network and systematically remove a set of users from the largest community.
We do so separately for the non-bipartisan nodes and bipartisan nodes.
We start by removing nodes in ascending order of degree, in deciles upon each iteration. This process is repeated for ten iterations for each group until all the nodes are removed. 
Upon each iteration, we check whether the retweet network contains more echo chambers based on the homophily of communities. 
Specifically, we assess whether a new community emerges as separate from the existing community and whether this new community exhibits polarity close to 1 and -1.
We use the same community detection approach (Louvain method with the same resolution) and echo chamber analysis as in Section~\ref{sec: echo chamber}.  

\begin{figure}[t!]
    \centering
    \subfloat[\textit{The number of community}\label{fig:remove_degree_numberofcommunity}]{
    \includegraphics[width = 0.23\textwidth]{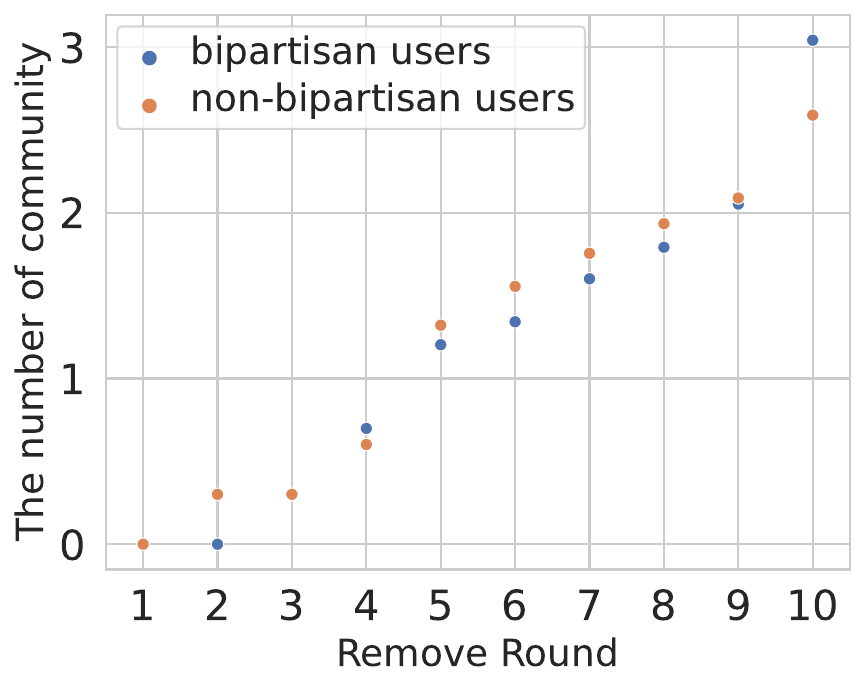}
    }
    \subfloat[\textit{The distribution of community polarity}\label{fig:remove_dgree_box}]{
        \includegraphics[width = 0.23\textwidth]{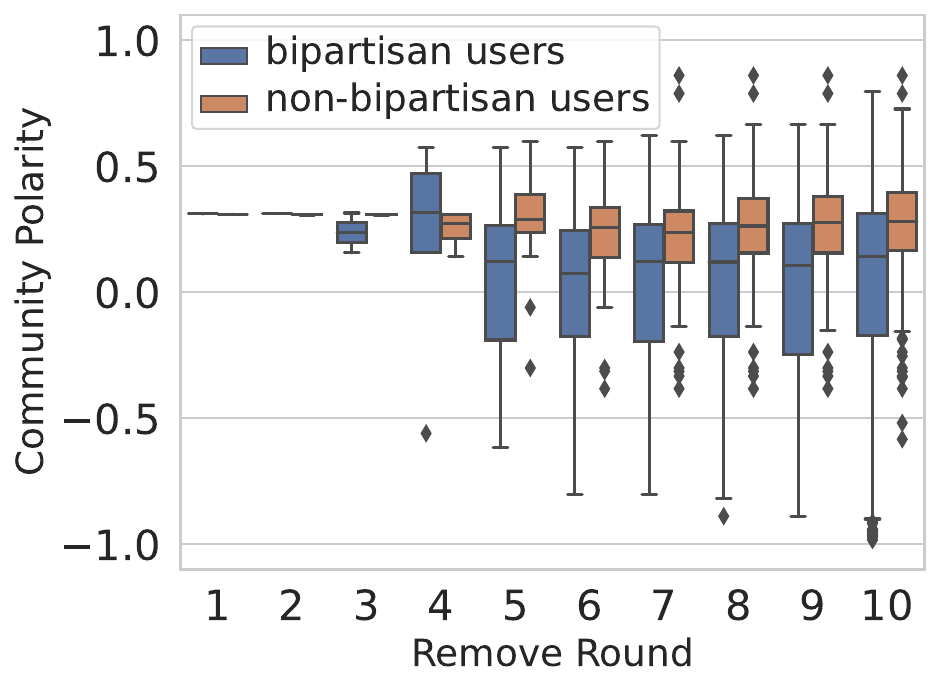}
        }        
        \caption{The result of the nodes removing with 10\% of the nodes being removed each time in ascending order of degree are considered from the number of communities and the distribution of the community polarity. The $x$-axis of~\ref{fig:remove_degree} and ~\ref{fig:remove_degree_numberofcommunity} represent the round of removal. The $y$-axis of~\ref{fig:remove_degree_numberofcommunity} represents, on a logarithmic scale base 10 quantifies the number of communities at each removal stage. }
    \label{fig:remove_degree}
\end{figure}


\pb{Results:} 
Figure~\ref{fig:remove_degree} presents the progressive outcomes following each round of node removal. 
Figure~\ref{fig:remove_degree_numberofcommunity} displays the number of communities in the network after each removal step. 
The $y$-axis represents, on a logarithmic scale base 10, quantifies the number of communities at each removal stage. 
Figure~\ref{fig:remove_dgree_box} also shows the distribution of community polarity resulting from the removal of nodes.
As nodes are removed from the graph, the communities become more fragmented. 
Following the removal of bipartisan nodes, the largest community contains 92.93\%  of the total nodes, and the remaining nodes are divided into 1,229 communities, with a proportion of 98.86\% consisting of single-node communities. 
After the removal of non-bipartisan nodes, the largest community contains 93.01\% of nodes and the remaining nodes are partitioned into 404 communities, with a proportion of 99.75\% communities only having one node.
After removing the bipartisan nodes, the resulting communities show a clearer trend, including more polarised communities.
This results in those smaller communities being identified.
Following the removal of all bipartisan nodes, the resulting smaller communities consist of 30.68\% \textit{pro Russia} users, 57.13\% \textit{pro Ukraine} users, and the remaining portion comprising \textit{not sure} users.
However, the result of the non-bipartisan nodes, the resulting smaller communities consist of 5.40\% \textit{pro Russia} users, 66.32\% \textit{pro Ukraine} users, 6.94\% \textit{bipartisan users}, and the remaining portion comprising \textit{not sure} users.
The removal of bipartisan nodes leads to increased fragmentation of \textit{pro Russia} communities, as more \textit{pro Russia} nodes form new communities. 



We take the polarity of communities as evidence of echo chambers, characterised by frequent interactions within the community and the sharing of similar polarity in Section~\ref{sec: echo chamber}.
The above results show that the presence of bipartisan nodes plays a key role in preventing the emergence of polarised communities.
Removing bipartisan nodes leads to an increase in the number of communities exhibiting polarity close to 1 and -1.
We argue that the bipartisan nodes, therefore, induce the number of polarised communities, indicating that the existence of \textit{bipartisan users} decreases the prominence of echo chambers.
\section{Conclusion} \label{sec: conclusion}

In this work, we have explored the presence of \textit{bipartisan users} and echo chambers in online discussions related to the Russo-Ukrainian war. 
Our analysis reveals evidence of echo chambers in the retweet network. 
Users share similar polarity with their neighbours in the retweet network corresponding to \textit{pro Russia} and \textit{pro Ukraine} groups. 
Additionally, most communities exhibit clear polarity and some small communities show extreme polarity.
We also detected a group of users who are bipartisan.
Such users share information showing both \textit{pro-Russia} and \textit{pro-Ukraine} polarity.
We checked whether their bipartisan attitude damages their influence in the discussion and how they interact with others. 
The result shows that the retweet counts do not benefit partisan groups or \textit{bipartisan users}.
However, for the like influence, \textit{pro Russia} group benefits from their partisan and \textit{pro Ukraine} group does not. 
Finally, we investigated whether these users might mitigate the echo chamber effect. 
The result shows that, compared with the control group, removing bipartisan nodes from the retweet network leads to more communities with extreme polarity, indicating that the bipartisan nodes connect the clusters of consistent users together to decrease the echo chamber.

\textbf{Limitations and Future Work:}
A key limitation of our analysis is the inherent limitations of the classifier used to categorise tweets, as there is room for improvement in performance. We note the difficulty of underlying tasks due to the variety of entities and the respective stances involved. Albeit, we train 30 different models with a grid search for hyper-parameters to achieve the best performance for each combination of embedding and algorithm.
The analysis of the bipartisan group is further limited to a two-week time period. 
We conjecture that a longer timeframe may expose greater bipartisanship, as users may vary their opinions with time. 
A key line of future work is exploring more longitudinal patterns.


\bibliographystyle{IEEEtran}
\bibliography{aaai22}

\end{document}